\documentclass[final,5p,times,twocolumn,sort&compress]{elsarticle}

\usepackage{epsfig}
\usepackage{amsmath}
\usepackage{url}

\usepackage{graphicx}
\usepackage{lineno}
\usepackage{color}
\usepackage{ulem}

\journal{Physics Letter B}

\begin{document}

\begin{frontmatter}

\title{Spectroscopic factor and proton formation probability for the $d_{3/2}$ proton emitter $^{151m}$Lu}

\author[bh]{F. ~Wang}
\author[bh]{B. H. ~Sun\corref{cor1}}
\ead{bhsun@buaa.edu.cn}
\author[imp,surrey]{Z. ~Liu\corref{cor1}}
\ead{liuzhong@impcas.ac.cn}
\author[lvp]{R. D. ~Page}
\author[kth]{C. ~Qi}
\author[jyu]{C. ~Scholey}
\author[surrey]{S. F. ~Ashley}
\author[lvp]{L. ~Bianco}
\author[surrey]{I. J. ~Cullen}
\author[iaea]{I. G. ~Darby}
\author[jyu]{S. ~Eeckhaudt}
\author[triumf]{A. B. ~Garnsworthy}
\author[surrey]{W. ~Gelletly}
\author[upc]{M. B. ~Gomez-Hornillos}
\author[jyu]{T. ~Grahn}
\author[jyu]{P. T. ~Greenlees}
\author[york]{D. G. ~Jenkins}
\author[surrey]{G. A. ~Jones}
\author[ithemba]{P. ~Jones}
\author[lvp]{D. T. ~Joss}
\author[jyu]{R. ~Julin}
\author[jyu]{S. ~Juutinen}
\author[jyu]{S. ~Ketelhut}
\author[um]{S. ~Khan}
\author[um]{A. ~Kishada}
\author[jyu]{M. ~Leino}
\author[tokyo]{M. ~Niikura}
\author[irmm]{M. ~Nyman}
\author[lvp]{J. ~Pakarinen}
\author[gsi]{S. ~Pietri}
\author[surrey]{Z. ~Podolyak}
\author[jyu]{P. ~Rahkila}
\author[lvp]{S. ~Rigby}
\author[jyu]{J. ~Saren}
\author[jaea]{T. ~Shizuma}
\author[jyu]{J. ~Sorri}
\author[surrey]{S. ~Steer}
\author[lvp]{J. ~Thomson}
\author[surrey]{N. J. ~Thompson}
\author[jyu]{J. ~Uusitalo}
\author[surrey]{P. M. ~Walker}
\author[ua]{S. ~Williams}
\author[lzu]{H. F. ~Zhang}
\author[imp]{W. Q. ~Zhang}
\author[bh]{L. H. ~Zhu}


\cortext[cor1]{Corresponding Author}
\address[bh]{School of Physics and Nuclear Energy Engineering, Beihang University, Beijing 100191, China}
\address[imp]{Institute of Modern Physics, Chinese Academy of Sciences, Lanzhou 730000, China}
\address[surrey]{Department of Physics, University of Surrey, Guildford, Surrey GU2 7XH, UK}
\address[lvp]{Department of Physics, Oliver Lodge Laboratory, University of Liverpool, Liverpool L69 7ZE, UK}
\address[kth]{KTH, Alba Nova University Center, SE-10691 Stockholm, Sweden}
\address[jyu]{University of Jyvaskyla, Department of Physics, P.O. Box 35, FI-40014 University of Jyvaskyla, Finland}
\address[iaea]{Department of Nuclear Sciences and Applications, International Atomic Energy Agency, A-1400 Vienna, Austria}
\address[triumf]{TRIUMF, 4004 Wesbrook Mall, Vancouver, British Columbia V6T 2A3, Canada}
\address[upc]{Universitat Polit\'{e}cnica de Catalunya (UPC), 08034 Barcelona, Spain}
\address[york]{Department of Physics, University of York, Heslington, York, UK YO10 5DD, UK}
\address[ithemba]{iThemba LABS, National ResearchFoundation, PO Box 722, Somerset West, South Africa}
\address[um]{Schuster Building, School of Physics and Astronomy, University of Manchester, Manchester M13 9PL, UK}
\address[tokyo]{CNS, University of Tokyo, Tokyo 351-0100, Japan}
\address[irmm]{European Commission, Joint Research Centre, Institute for Reference Materials and Measurements, Retieseweg 111, B-2440 Geel, Belgium}
\address[gsi]{GSI Helmholtzzentrum f\"{u}r Schwerionenforschung, D-64291 Darmstadt, Germany}
\address[jaea]{Japan Atomic Energy Agency, Tokai, Ibaraki 319-1195, Japan}
\address[ua]{Nikhef National Institute for Subatomic Physics, University of Amsterdam, Amsterdam, Netherlands}
\address[lzu]{School of Nuclear Science and Technology, Lanzhou University, Lanzhou 730000, China}

\begin{abstract}
The quenching of the experimental spectroscopic factor for proton emission from the short-lived $d_{3/2}$ isomeric state in $^{151m}$Lu was a long-standing problem. In the present work, proton emission from this isomer has been reinvestigated in an experiment at the Accelerator Laboratory of the University of Jyv\"{a}skyl\"{a}. The proton-decay energy and half-life of this isomer were measured to be 1295(5) keV and 15.4(8) $\mu$s, respectively, in agreement with another recent study. These new experimental data can resolve the discrepancy in the spectroscopic factor calculated using the spherical WKB approximation. Using the R-matrix approach it is found that the proton formation probability indicates no significant hindrance for the proton decay of $^{151m}$Lu.
\end{abstract}

\end{frontmatter}



\section{Introduction}
\label{Section:Intro}

Proton emission is a quantum tunneling process in which the escaping proton penetrates through a potential barrier consisting of Coulomb and centrifugal potentials.
The study of proton decay provides critical spectroscopic information on the proton emitters and the ordering of quantum states of nuclei lying beyond the proton drip line~\cite{woods1997,blank2008nuclear,pfutzner2012radioactive,Delion2006113}. The spectroscopic factor is conventionally employed as a measure of the purity of the single-particle configuration of the initial wave function.

The experimental spectroscopic factor ($S_p^{\rm exp}$) is usually defined as the ratio between the experimental half-life and the calculated one based on single-particle models. It provides a measure of the amplitude of the single particle ($n, l, j$) component in the proton-emitting nucleus. The calculated proton half-life $t_{1/2}^p$(calc) can be obtained using the WKB approximation and has a very strong dependence on the proton-decay energy, the orbital angular momentum carried by the emitting proton as well as the effective single-particle potential and the corresponding initial single-proton wave function used in the calculation. One can assert that the way one extracts the experimental spectroscopic factor is an effective theory since one has to introduce an effective single-proton potential to mimic the motion of the decaying proton inside the nucleus. The calculated penetration probability and the extracted experimental spectroscopic factor are sensitive to that potential, as already indicated in various calculations~\cite{PhysRevC.55.2255,PhysRevC.56.1762,PhysRevC.68.054301,ZhangHF-JPG2010,Dong2009PhysRevC.79.054330,Bhattacharya2007263,Ferreira2011508,Long2014PhysRevC.90.054326,Xia2016PhysRevC.93.014314,Qian2016}.

The experimental spectroscopic factor ($S_p^{\rm exp}$) may be compared with the theoretical one $S_p^{\rm th}$.
The latter is model dependent and very sensitive to the nuclear structure involved,  including the single-particle energies, which are much affected
by the nuclear potential used and the excitation modes.
Within the BCS theory the spectroscopic factor is given by $S_p^{\rm th} = u^2_j$, where the vacancy factor $u^2$ is the probability that the spherical
shell-model orbital with ($n, l, j$) quantum numbers is empty in the daughter nucleus.
The agreement between experimental and theoretical spectroscopic factors can be a good indication that
a reasonable and consistent initial wave function for the outgoing proton has been taken.
Considering the large uncertainties mentioned above, however, one may not be able to draw a firm conclusion.

Proton emitters in the region with $A \approx 150-170$,  $69\le Z \le 79$ are spherical or nearly spherical. They are of particular interest
as the $s_{1/2}$, $d_{3/2}$, and $h_{11/2}$ proton orbitals are almost degenerate. This leads to the presence of low-spin and high-spin states in close proximity.
Systematic analysis of the experimental data~\cite{PhysRevC.55.2255,PhysRevC.56.1762,PhysRevC.68.054301,
ZhangHF-JPG2010},
shows good agreement of the \mbox{theoretical} spectroscopic factors with the experimental ones for $h_{11/2}$ and $s_{1/2}$ emitters.
In contrast, for $d_{3/2}$ states the observed spectroscopic factors are systematically lower than those predicted by, e.g.,
 a low-seniority spherical shell model calculation~\cite{PhysRevC.55.2255} or BCS calculations~\cite{PhysRevC.56.1762}.

In order to address the discrepancies between experimental and theoretical spectroscopic factors, sophisticated models have been
developed to evaluate the role of dynamical particle-vibration coupling~\cite{PhysRevC.64.034317,PhysRevC.64.041304},
or the effect of non-negligible deformation.

More recent calculations of the spectroscopic factors, e.g., within a generalized liquid drop model~\cite{Dong2009PhysRevC.79.054330},
using the covariant density functional theory~\cite{Bhattacharya2007263,Ferreira2011508,Long2014PhysRevC.90.054326,Xia2016PhysRevC.93.014314}
or a deformed density-dependent model~\cite{Qian2016},
do not show the apparent systematic trends as predicted by the low-seniority shell model~\cite{PhysRevC.55.2255} or BCS calculations~\cite{PhysRevC.56.1762}.

An alternative description of the proton-decay process
is given by the R-matrix approach~\cite{Delion2006113}. It provides a microscopic scheme to extract the experimental
proton formation amplitude at the nuclear surface in a model independent way~\cite{Qi2012-PhysRevC.85.011303}. In this scheme,
as will be illustrated in the Discussion Section, the proton decay process can be evaluated in two steps:
the inner process which describes the dynamic motion of the proton inside the nucleus and the possibility for it to be emitted,
and the outer process which describes the penetration of the proton through the barrier. The latter part of the inner process corresponds to
the proton formation amplitude that reflects the overlap between the parent and daughter wave functions, from which one can distinguish
the role played by deformation and pairing on the decay process. This scheme avoids the ambiguities of the deduced spectroscopic factor
in relation to the surface effects and quantifies in a more precise manner the nuclear many-body structure effects. It
is also valid for all charged particle decays. It is worth noting that, if a smooth effective single-proton potential is used in calculating
the spectroscopic factor, the proton formation amplitude and the effective spectroscopic factor may show a similar systematic pattern.

An important case is $^{151m}$Lu~\cite{prc1999}, the heaviest odd-A nuclide for which proton emission has been observed from a $d_{3/2}$ isomeric state.
However, the observed half-life was much longer than that predicted by spherical WKB calculations and the extracted $S^{exp}_{p}$ of 0.26$^{+0.14}_{-0.08}$
using the WKB approximation~\cite{prc1999} is lower than the predicted values of $S^{th}_{p}$ of 0.73~\cite{Cwiok1987379} or 0.67~\cite{PhysRevC.55.2255}.
However, a recent study by Taylor et al.~\cite{PhysRevC.91.044322} reported a lower value for the proton-decay energy, which could potentially
resolve the discrepancy. In that work, nonadiabatic quasiparticle calculations were also performed, which were able to reproduce the
improved experimental data, provided that $^{151m}$Lu has a deformation of $\beta_2 \approx$ -0.12. This value is comparable to
the corresponding $\beta_2$ value deduced for the ground state (g.s.) of $^{151}$Lu using the same formalism~\cite{Procter2013Phys.Lett.B725}.
In addition, these calculations were able to reproduce properties of excited levels built upon the proton-emitting states.

Here we report on a reinvestigation of $^{151m}$Lu in an independent recoil-decay tagging (RDT) experiment performed at the University of Jyv\"{a}skyl\"{a}.
Our new results are consistent with those of Ref.~\cite{PhysRevC.91.044322} and we discuss the experimental and theoretical spectroscopic factor
for $^{151m}$Lu assuming a spherical shape with the WKB approximation~\cite{ZhangHF-JPG2010}. Considering that nuclear structure effects are
not included in the WKB barrier transmission approximation and the model dependence of theoretical spectroscopic factors, we introduce the proton
formation probability as a more proper description of the proton-decay process~\cite{Qi2012-PhysRevC.85.011303}, which defines the possibility
of finding the decaying proton at the nuclear surface. The proton formation probability extracted from the present results indicates no significant hindrance for the proton decay of $^{151m}$Lu.

\section{Experimental details and results}
\label{Section:ExpResults}

\begin{figure*}
\centering
\begin{minipage}{.65\textwidth}
\centering
  \includegraphics[width=11cm]{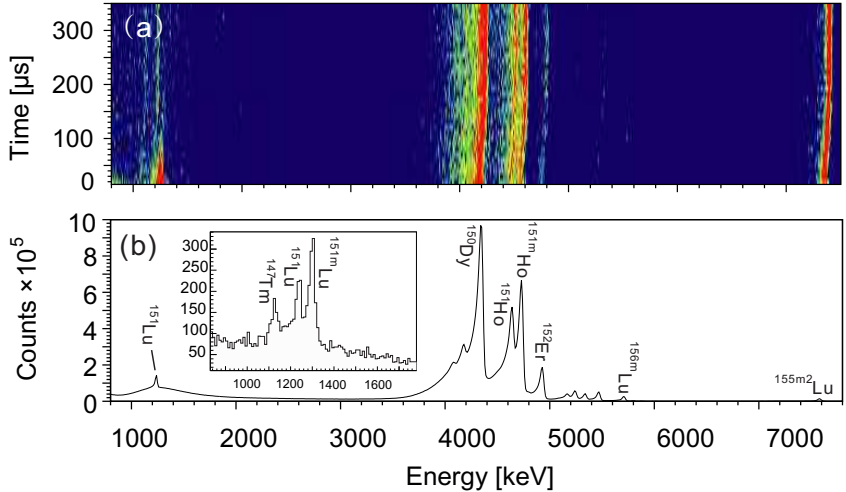}
\end{minipage}%
\begin{minipage}{.3\textwidth}
\centering
  \caption{(a) PIN-vetoed decay spectrum of DSSDs as a function of time after implantation. (b) Projection of the decay spectrum within 500 ms after the implantation. The inset is the projection between 30 - 400 $\mu$s,  where both proton lines from $^{151g,m}$Lu are clearly visible.
  }
\label{fig:exp}
\end{minipage}
\end{figure*}

The experimental setup consisted of the JUROGAM Ge-detector array~\cite{juro} at the target position,
the gas-filled recoil separator RITU~\cite{ritu,ritu1} and the GREAT spectrometer at the focal plane of RITU.
In this experiment, excited states of $^{151}$Lu were populated by bombarding a self-supporting 500 $\mu$g/cm$^2$ isotopically enriched $^{96}$Ru target
with a $^{58}$Ni beam at 266 MeV and 274 MeV delivered by the K130 cyclotron. A 50 $\mu$g/cm$^2$ C charge reset foil was placed behind the target.
The average beam current on the target was 3 particle nA for 110 hours.
After a time of flight of about 0.6 $\mu$s in RITU, the evaporation residues passed through a gas-filled multi-wire proportional chamber (MWPC),
and then were implanted into a pair of 300 $\mu$m thick double-sided silicon strip detectors (DSSDs) of the GREAT spectrometer. This spectrometer registers
the recoiling evaporation residues, proton and $\alpha$ decays, $\beta$ rays, conversion electrons as well as X and $\gamma$ rays.
Each DSSD is segmented into 40 horizontal strips in the front and 60 vertical strips at the back, providing a total of 4800 pixels. To minimize the interference
from scattered electrons and light ions in the DSSDs, a PIN-diode detector array surrounding the DSSDs in GREAT can be used as a veto.
Prompt $\gamma$ rays emitted in the fusion-evaporation reactions were detected by the JUROGAM array comprising 43 Ge detectors.
More details of the setup can be found in Refs.~\cite{ritu,great}.

For each event, all signals induced in the JUROGAM, MWPC and GREAT were recorded by a triggerless data acquisition system~\cite{tdr}.
In this system, all channels were running independently and each registered signal was time stamped by a 100 MHz clock. Thus the prompt $\gamma$ rays
at the target position, the impinging time and position of the evaporation residues, as well as the energy, time and position of subsequent decays, could be measured and stored.
A position-energy-time-correlation analysis of the event chains allowed one to make detailed deductions
with implantation rates of several hundred evaporation residues per second.
In other words, decays within a given pixel of DSSDs can be correlated with the previous implant in the same pixel.
In this way it is possible to determine the decay time of the radioactivity.

In total, about 1500 full-energy protons were registered for the $d_{3/2}$ isomeric decays, 80 percent of which are from the setting with a beam energy of 266 MeV.
Half-lives in the range of microseconds to about a few hundreds of milliseconds could be measured by observing the decay of the activity. The data were analyzed with the
GRAIN~\cite{grain} and RADWARE~\cite{radware} software packages.

The energy-time spectrum of the charged particle decay is shown in Fig.~\ref{fig:exp}(a) with a bin size of 10 $\mu$s in time.
The PIN-diode detector array surrounding the DSSDs in GREAT were used as a veto.
A few $\alpha$ particle peaks are clearly resolved. It can be seen that the peak energies increase with
time in the first 200 $\mu$s after the implantation,
and then remain constant for decays thereafter, causing the energy resolution to be degraded for fast decays.
This is due to the residual pulse height in the associated amplifiers caused by the implant at the time of decay.
Such energy resolution degradation was also observed, for example, in Refs.~\cite{Batchelder1998PhysRevC.57.R1042,prc1999},
and the correction is necessary for life times up to a few milliseconds.

The energy projection of the decay spectrum within 500 ms after implantation is presented in Fig.~\ref{fig:exp}(b). One proton peak is visible in the low-energy region.
Peaks above 4 MeV are \mbox{assigned} to known $\alpha$-decaying nuclei. The most intense $\alpha$-particle peaks between 4 and 5 MeV are from the decays of
the $N=84$ isotones ($^{150}$Dy, $^{151}$Ho, $^{152}$Er). The higher-energy $\alpha$ lines, including the isomeric transition of $^{155m2}$Lu, are
due to the isotopic impurities of heavier Ru isotopes in the target.

The kinetic energies of these $\alpha$ particles and the proton from the g.s. of $^{151}$Lu ~\cite{zpa1982} were used for the energy calibration of the DSSDs.
Corrections~\cite{HofmannZPA1979,Page1996PhysRevC.53.660} were applied to take into account the pulse height defect for protons and $\alpha$ particles in silicon~\cite{LENNARD1990NIMB281}, the contribution of the recoiling daughter nucleus to the energy signal~\cite{RATKOWSKI1975NIM533} and the non-linear response of silicon detectors for low-$Z$ ions~\cite{LENNARD1986NIMA454,LENNARD1990NIMB47}.
As shown in the inset in Fig.~\ref{fig:exp}(b), the $^{151m}$Lu proton-decay peak is clearly resolved from the g.s. proton-decay line with an energy difference of 62 keV, producing an $E_p$ of 1295(5) keV for the isomeric state.
The new value is consistent with 1310(10) keV obtained in Ref.~\cite{prc1999}
and the lower value of 1285(4) keV reported recently in Ref.~\cite{PhysRevC.91.044322}.  The corresponding proton-decay energy $Q_p$ was calculated to be 1317(5) keV taking into account the recoiling energy of the daughter nucleus and (electron) screening correction~\cite{HUANG1976243}.
The proton-decay half-life associated with the isomer is $15.4\pm 0.8$ $\mu$s,
which compares with the previously reported values of $16(1)$ $\mu$s~\cite{prc1999} and $17(1)$ $\mu$s~\cite{PhysRevC.91.044322}.

\begin{figure}[htb!]
  \centering
  \includegraphics[width=0.5\textwidth]{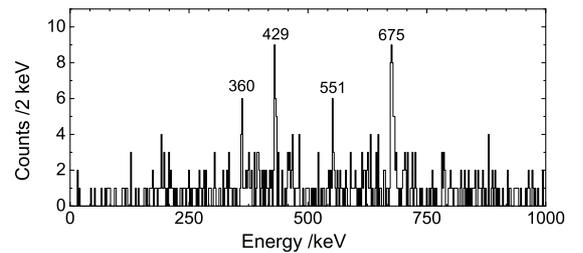}\\
  \caption{Prompt $\gamma$-ray spectrum tagged with proton decay from the $^{151m}$Lu.
}\label{fig:fig2}
\end{figure}

By tagging on the protons emitted from $^{151m}$Lu,  four prompt $\gamma$-ray transitions feeding the $d_{3/2}$ isomer are identified at 675, 551, 429 and 360 keV as shown in Fig.~\ref{fig:fig2}. Here the correlation time between the proton decay and the previously implanted recoil is within 80 $\mu$s.
The first three of these were also observed in Ref.~\cite{PhysRevC.91.044322}, but the 369 and 283 keV transitions reported there are not confirmed in this work. The relative intensities of these four transitions are also listed in Table~\ref{tbl}.  Although the 675 keV transition is the strongest $\gamma$ ray in the spectrum, no
 $\gamma$-$\gamma$ coincidences could be established due to the low statistics.
Consequently the level scheme proposed in Ref.~\cite{PhysRevC.91.044322} cannot be confirmed. It should be pointed out that if combining Fig.2 and the corresponding Fig.3(b) in~\cite{PhysRevC.91.044322} the 675 keV $\gamma$ ray will be the strongest one feeding the $d_{3/2}$ isomer.

\begin{table}
\begin{center}
\caption{Energies and efficiency-corrected relative intensities for $\gamma$-ray transitions assigned to $^{151m}$Lu. The relative intensities are normalized to that of the 675 keV transition.
}\label{tbl}
\begin{tabular}{cc}
\hline \hline
$E_{\gamma}/keV$ &$I_{\gamma}$    \\
\hline
675 &100(28) \\
429 &35(14) \\
360 &33(15) \\
551 &30(15) \\
\hline \hline
\end{tabular}
\end{center}
\end{table}

\section{Discussion}
\label{Section:Discussion}

\begin{figure} [htp!]
  \centering
  \includegraphics[width=0.45\textwidth]{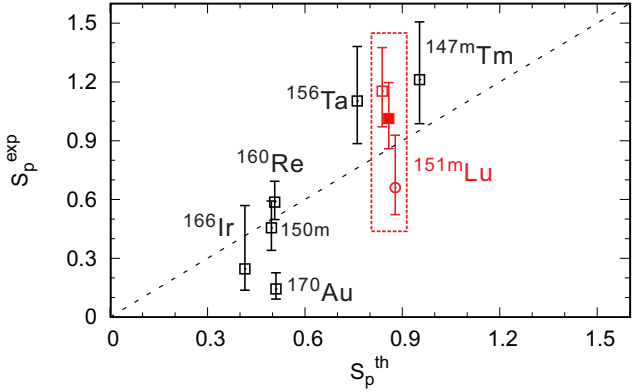}\\
  \caption{Experimental spectroscopic factors vs. theoretical ones obtained in the RMF+BCS theory for the region with $64 \le Z \le 82$ for $d_{3/2}$ states. The results of $^{151m}$Lu (in red color) deduced from the present work, Refs.~\cite{prc1999} and ~\cite{PhysRevC.91.044322}, are indicated by the solid square,  open circle and open square, respectively. The $S_p^{{\rm th}}$ of $^{151m}$Lu are shifted slightly for a better view. The symbol ($m$) denotes an isomeric state. }\label{fig:d32}
\end{figure}

The slight decrease in $Q_p$
compared with the original value~\cite{prc1999} increases the
theoretical proton partial half-life within the WKB approximation.
To illustrate the effect of the present data, we have calculated the theoretical spectroscopic factor using the relativistic mean field theory (RMF) combined with the BCS method as described in Ref.~\cite{ZhangHF-JPG2010}.
The experimental and theoretical spectroscopic factors are plotted in Fig.~\ref{fig:d32} for the $d_{3/2}$ proton
emitters in the subshell between $Z$ = 64 and 82.
The error bars have taken into account the experimental errors
 of the half-lives and the theoretical errors originated from the uncertainties in $Q_p$.
With the data obtained in the present work, the experimental spectroscopic factor $S_p^{\rm exp}$ of $^{151m}$Lu increase from 0.66~\cite{prc1999} to
1.01$^{+0.18}_{-0.15}$. This value should be compared with $S_p^{\rm th}$ of 0.86.
Thus the present results produce a larger experimental spectroscopic factor $S_p^{\rm exp}$ and the spectroscopic factor discrepancy in~\cite{prc1999} can be resolved.
For comparisons, the corresponding spectroscopic factors recalculated based on Refs.~\cite{PhysRevC.91.044322} and~\cite{prc1999}, are also shown in Fig.~\ref{fig:d32}. Our new result is somewhat in between.

\begin{figure}[htbp]
  \centering
  \includegraphics[width=0.4\textwidth]{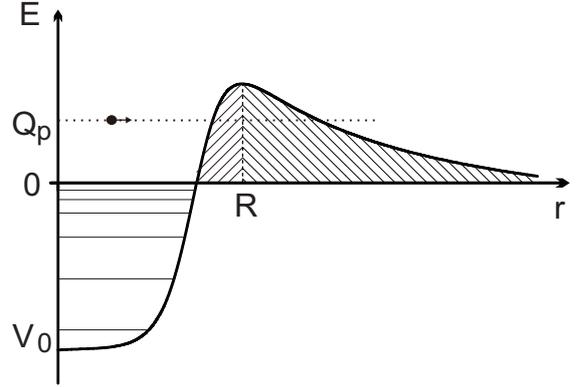}
\caption{Schematic view of the proton emission process in the R-matrix approach with no contribution from the centrifugal barrier (i.e.  zero orbital angular momentum).
For details please refer to the text. }
\label{fig:decayscheme}
\end{figure}

However, the calculated spectroscopic factor is rather interaction or model dependent as mentioned in the introduction, so it is not an ``ideal" quantity to describe the proton-decay process.
Shown in Fig.~\ref{fig:decayscheme} is a schematic plot for the R-matrix description of the proton emission process, where the decay process is divided into two regions: the inner and outer region.
Here the value $R$ defines a radius of the nuclear surface outside of which the nuclear potential vanishes.
In the outer region,  the nuclear attraction vanishes and the proton escapes the nucleus with a rate solely determined by the $Q_p$ value, the Coulomb and centrifugal barriers. In the inner region,  the dynamic motion of the proton determines the proton-decay formation property that describes the influence of nuclear structure on the proton decay~\cite{Qi2012-PhysRevC.85.011303}.
This can be contrasted with the WKB calculations of the total penetrability that are also influenced by the surface part and the inner part through the effective single-proton potential.

As shown in Fig.~2 in Ref.~\cite{Qi2012-PhysRevC.85.011303}, the formation amplitudes log$_{10}(|$RF(R)$|^2)$ extracted from experimental data,
are clearly divided into two distinct groups characterized by deformation: the left region for the decays of the well prolate-deformed nuclei for lighter isotopes and
the right region for the decays of spherical and oblate-deformed nuclei for heavy isotopes.
The proton decay of $^{151m}$Lu (located at $\rho'\approx 21$ and log$_{10}|$RF(R)$|^2 \approx$ -1.5 in Fig.~2 in Ref.~\cite{Qi2012-PhysRevC.85.011303})
falls basically in the spherical and oblate-deformed group but is slightly below the overall trend.

\begin{figure}
  \centering
  \includegraphics[width=0.45\textwidth]{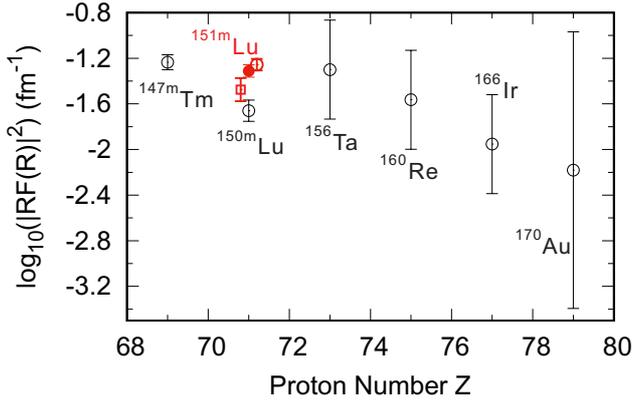} 
  \caption{Proton-decay formation amplitudes log$_{10}|$RF(R)$|^2$ extracted from experimental data as a function of $Z$ for the $d_{3/2}$ states.
  The results of $^{151m}$Lu (in red color) deduced from the present work, Refs.~\cite{prc1999} and ~\cite{PhysRevC.91.044322}, are indicated by the solid circle, open square and open circle, respectively.  For display, the previous results are shifted slightly in $Z$.
}\label{fig:log10}
\end{figure}

Using the data obtained in the present work the proton formation probability $|RF_l(R)|^2$ of $^{151m}$Lu is recalculated to be 0.049(3)~fm$^{-1}$. This is 48\% larger than that obtained from the adopted data~\cite{blank2008nuclear} and is nearly identical to those for the $d_{3/2}$ \mbox{states} in neighboring nuclei, e.g., $^{147m}$Tm and $^{156}$Ta~\cite{Qi2012-PhysRevC.85.011303}.
The proton formation probabilities for the six $d_{3/2}$ proton emitters are shown in Fig.~\ref{fig:log10}.
The new proton formation probability for $^{151m}$Lu fits well in the group of spherical and oblate-deformed proton emitters.

The 675-keV transition in Fig.~2 is the most intense one feeding the $d_{3/2}$ isomer, in agreement with the observation in Ref.~\cite{PhysRevC.91.044322}. Thus the 7/2$^+$ $\rightarrow$ 3/2$^+$ transition could be 675 keV instead of 546 keV as assumed in Ref.~\cite{PhysRevC.91.044322}, then the nonadiabatic quasiparticle calculations (see Fig.~6 in Ref.~\cite{PhysRevC.91.044322}) would interpret the quadrupole deformation $\beta_2$ as around -0.07.
Furthermore, a slightly reduced half-life of the 3/2$^+$ isomeric state (from 16.5(7) $\mu$s to 15.4(8) $\mu$s), will also give a smaller $\beta_2$ value (see Fig.~7 in Ref.~\cite{PhysRevC.91.044322}). Considering the theoretical uncertainty, this would be consistent with the spherical nature suggested by the proton formation probability in the present work.
It is noticed in Fig.~\ref{fig:log10} that the formation probability for the $d_{3/2}$ state in $^{150m}$Lu is still obviously lower than those for the neighboring nuclei. The reason is unclear and a new precision measurement may be necessary to clarify the situation.

\section{Summary}
\label{Section:Summary}

The $d_{3/2}$  proton emitter $^{151m}$Lu has been reinvestigated using the RDT technique. With the decay energy and half-life values measured in the present work
the spectroscopic factor for $^{151m}$Lu is increased from 0.66 to about 1 in the WKB approach, solving the long-standing spectroscopic factor quenching problem
in Ref.~\cite{prc1999}. The decay of the $d_{3/2}$ proton emitters was also discussed in terms of the proton formation probability,
a more proper and microscopic quantity to describe the proton decay process. The extracted proton formation probability for $^{151m}$Lu is
compared to those in neighboring nuclei,and is found to follow well the general trend of spherical proton emitters.

\section*{\uppercase{acknowledgments}}
\label{acknowledgments}
This work has been supported by the National Natural Science Foundation of China under
Nos. 11475014, 11435014, 11405224, 11205208, 11675225, 11675066, U1632144 and the National key research and development
program (2016YFA0400500), the ``100 Talented Project" of the Chinese Academy of Sciences,
the EU 6th Framework programme ``Integrating
Infrastructure Initiative - Transnational
Access", \mbox{Contract} Number: 506065 (EURONS) and by the Academy of
Finland under the Finnish Centre of Excellence Programme 2006-2011
(Nuclear and Accelerator Based Physics Programme at JYFL), and United Kingdom Science and Technology Facilities Council.
CS(209430) and PTG(111965) acknowledge the support of the Academy of Finland.
CQ is supported by the Swedish Research Council (VR) under grant Nos. 621-2012-3805, and
621-2013-4323 and the G\"oran Gustafsson foundation. The computations were performed on resources provided by the Swedish National Infrastructure for Computing (SNIC) at
PDC, KTH, Stockholm. We would also like to thank Bao-An Bian, Cary Davids, Lang Liu, and Zheng-Hua Zhang for valuable discussions.


\end{document}